\documentclass{appolb}
\usepackage{graphicx}
\usepackage{amsmath}
\newcommand{\tr}{\mathop{\mathrm{Tr}}}
\newcommand{\slparc}{\mbox{$\,\slash$ \hspace{-0.9em}$\partial$}}

\begin{document}
\title{Chiral phase transition in an extended linear sigma model:
initial results
\thanks{Presented at the Workshop on Unquenched Hadron Spectroscopy:
     Non-Perturbative Models and Methods of QCD vs. Experiment,\\ 
               At the occasion of Eef van Beveren's 70th birthday}%
}
\author{Gy. Wolf, P. Kov\'acs
\address{Institute for Particle and Nuclear Physics, Wigner Research
  Center for Physics, Hungarian Academy of Sciences, POB 49, H-1525 Budapest,
  Hungary}
\\[0.4cm]
Zs. Sz{\'e}p
\address{MTA-ELTE  Statistical and Biological Physics Research Group, H-1117 Budapest,
  Hungary}
}
\maketitle
\begin{abstract}
We investigate the scalar meson mass dependence on the chiral phase
transition in the framework of an SU(3), (axial)vector meson extended linear 
sigma model with additional constituent quarks and Polyakov loops. We 
determine the parameters of the Lagrangian at zero temperature in a hybrid 
approach, where we treat the mesons at tree-level, while the constituent 
quarks at 1-loop level. We assume two nonzero scalar condensates and together 
with the Polyakov-loop variables we determine their temperature dependence 
according to the 1-loop level field equations.
\end{abstract}
  
\section{Introduction}
The investigation of the QCD phase diagram is a very important subject both 
theoretically and experimentally nowadays. The ongoing and future heavy ion 
experiments such as RHIC, and CERN/LHC study the low density part of the phase 
diagram which can also be investigated theoretically by lattice QCD, at 
CBM/FAIR the high density part will be studied, which is still not settled 
theoretically, so it is worth to investigate the phase diagram thoroughly.

Our starting point is the (axial)vector meson extended linear sigma model with 
additional constituent quarks and Polyakov-loop variables. The previous 
version of the model, without constituent quarks and Polyakov-loops, was 
exhaustively analyzed at zero temperature in 
\cite{Parganlija_2013,KovacsWolf_anomaly_2013,KLVWZ_baryon_2014}\footnote{In the present work we use a different anomaly term ($c_1$ term). This, however, 
does not influence the results much.}. The Lagrangian of the model is given by,
\begin{align}
  \mathcal{L} & = \tr[(D_{\mu}\Phi)^{\dagger}(D_{\mu}\Phi)]-m_{0}%
  ^{2}\tr(\Phi^{\dagger}\Phi)-\lambda_{1}[\tr(\Phi^{\dagger}%
  \Phi)]^{2}-\lambda_{2}\tr(\Phi^{\dagger}\Phi)^{2}\nonumber \\
  & -\frac{1}{4}\tr(L_{\mu\nu}^{2}+R_{\mu\nu}^{2})+\tr\left[ \left(
      \frac{m_{1}^{2}}{2}+\Delta\right)
    (L_{\mu}^{2}+R_{\mu}^{2})\right]
  +\tr[H(\Phi+\Phi^{\dagger})]\nonumber \\
  & +c_{1}(\det\Phi+\det\Phi^{\dagger})+i\frac{g_{2}}{2}(\tr
  \{L_{\mu\nu}[L^{\mu},L^{\nu}]\}+\tr\{R_{\mu\nu}[R^{\mu},R^{\nu
  }]\})\nonumber \\
  & +\frac{h_{1}}{2}\tr(\Phi^{\dagger}\Phi)\tr(L_{\mu}
  ^{2}+R_{\mu}^{2})+h_{2}\tr[(L_{\mu}\Phi)^{2}+(\Phi R_{\mu}
  )^{2}]+2h_{3}\tr(L_{\mu}\Phi R^{\mu}\Phi^{\dagger})\\
  & +g_{3}[\tr(L_{\mu}L_{\nu}L^{\mu}L^{\nu})+\tr(R_{\mu}R_{\nu
  }R^{\mu}R^{\nu})] + g_{4}[\tr\left( L_{\mu}L^{\mu}L_{\nu}L^{\nu
    }\right) \nonumber\\
    & + \tr\left( R_{\mu}R^{\mu}R_{\nu}R^{\nu}\right)] + g_{5}\tr\left( L_{\mu}L^{\mu}\right) \,\tr\left(
    R_{\nu}R^{\nu}\right) + g_{6} [\tr(L_{\mu}L^{\mu})\,\tr(L_{\nu}L^{\nu})\nonumber \\
    & + \tr(R_{\mu}R^{\mu})\,\tr(R_{\nu}R^{\nu})] + \bar{\Psi}i
      \slparc\Psi - g_{F}\bar{\Psi}\left(\Phi_{S} +
        i\gamma_5\Phi_{PS}\right)\Psi,\nonumber 
\end{align}
where 
\begin{align}
D^{\mu}\Phi & = \partial^{\mu}\Phi-ig_{1}(L^{\mu}\Phi-\Phi
R^{\mu})-ieA_{e}^{\mu}[T_{3},\Phi] , \nonumber\\ 
L^{\mu\nu} & = \partial^{\mu}L^{\nu}-ieA_{e}^{\mu}[T_{3},L^{\nu}] -
\left\{\partial^{\nu}L^{\mu} - ieA_{e}^{\nu}[T_{3},L^{\mu}]\right\} , \nonumber \\
R^{\mu\nu} & = \partial^{\mu}R^{\nu} - ieA_{e}^{\mu}[T_{3},R^{\nu}]
- \left\{ \partial^{\nu}R^{\mu}-ieA_{e}^{\nu}[T_{3},R^{\mu}]\right\} . \nonumber
\end{align}
Here $\Phi$ stands for the scalar and pseudoscalar fields, $L^{\mu}$
and $R^{\mu}$ for the left and right handed vector fields, $\Psi=(u,
d, s)^{\text{T}}$ for the constituent quark fields, while $H$ for the
external field.

\section{Parametrization}
In order to go to finite temperature/chemical potential, parameters of
the Lagrangian have to be determined, which is done at $T=\mu=0$. For
this we calculate tree-level masses and decay widths of the model and
compare them with the experimental data taken from the PDG \cite{PDG}. 
For the comparison we use a $\chi^2$ minimalization method
\cite{MINUIT} to fit our parameters (for more details see
\cite{Parganlija_2013}). It is important to note that in the present
work we also included in the scalar and pseudoscalar masses the
contributions coming from the fermion vacuum fluctuations by adapting
the method of \cite{Chatterjee:2011jd}. 

We have $14$ unknown parameters, namely $m_0$, $\lambda_1$,
$\lambda_2$, $c_1$, $m_1$, $g_1$, $g_2$, $h_1$, $h_2$, $h_3$,
$\delta_S$, $\Phi_N$, $\Phi_S$, and $g_F$. Here $g_F$ is the coupling
of the additionally introduced Yukawa term, which can be determined
from the constituent quark masses through the equations $m_{u/d} =
g_F\phi_N/2$, $m_s = g_F\phi_s/\sqrt{2}$. 

It is worth to note that if we do not consider the very uncertain
scalar-isoscalar sector $m_0$, and $\lambda_1$ always appear in the
same combination $C_{1} = m_{0}^{2} + \lambda_{1} \left(\phi_{N}^{2}
  +\phi_S^2\right)$ in all the expressions, thus we can not determine
them separately. Additionally a similar combination appears for $m_1$
and $h_1$ in the vector sector as $C_{2} = m_{1}^{2} + \frac{h_{1}}{2}
\left(\phi_{N}^{2} + \phi_{S}^{2}\right)$ (see details in
\cite{Parganlija_2013}). The parameter values of the fit without scalars are 
given in Table~\ref{tab-param}.
\begin{table}
\centering
\caption{Parameters determined by $\chi^2$ minimalization}
\label{tab-param}   
\begin{tabular}{llll}
\hline
Parameter & Value & Parameter & Value \\\hline
$\phi_{N}$ [GeV]& $0.1622 $ & $h_{2}$ & $11.6586 $ \\
$\phi_{S}$ [GeV]& $0.1262 $ & $h_{3}$ & $4.7028 $ \\
$C_{1}$ [GeV$^2$] & $-0.7537 $ & $\delta_{S}$ [GeV$^2$] & $0.1534 $ \\
$C_{2}$ [GeV$^2$] & $0.3953 $ & $c_{1}$ [GeV]& $1.12 $ \\
$\lambda_{1}$ & undetermined &  $g_{1}$ & $-5.8943$ \\
$\lambda_{2}$ & $65.3221 $ & $g_{2}$ & $-2.9960$ \\
$h_{1}$ & undetermined & $g_{F}$ & $4.9429$ \\\hline
\end{tabular}
\end{table}
Since $\lambda_1$ and $h_{1}$ are undetermined they can be tuned to select the 
$f_0^L$ (a.k.a. $\sigma$) from the scalar spectrum (by its mass and decay 
widths) and its mass has, as we will see, a huge effect on the thermal 
properties of the model.

\section{Field equations}
\label{sec-eqn}

In our approach we have four order parameters, which are the $\phi_N$
non-strange and $\phi_S$ strange condensates, and the $\Phi$ and
$\bar{\Phi}$ Polyakov-loop variables. The condensates arise due to the
spontaneous symmetry breaking\footnote{Since isospin symmetry is
  assumed, we have only two condensates: $\phi_N$ and $\phi_S.$}, while
the Polyakov-loop variables naturally emerge in mean field approximation,
if one calculates free fermion grand canonical potential on a constant
gluon background. The effect of fermions propagating on a constant
gluon background in the temporal direction formally amounts to the
appearance of imaginary color dependent chemical potentials (for
details see \cite{Korthals_1999,Marko_2010}).

At finite temperature/baryochemical potential we can set up four
coupled field equations for the four fields, which are just the
requirements that the first derivatives of the grand canonical
potential according to the fields must vanish. As a first
approximation we apply a hybrid approach in which we only consider
vacuum and thermal fluctuations for the fermions, but not for the
bosons. 
We use a mean field Polyakov-loop potential $U(\Phi,\bar \Phi)$ of a polynomial 
form with coefficients determined in \cite{Ratti:2005jh}. Within this 
simplified treatment the equations are the following:
\begin{align}
-\frac{d }{d \Phi}\left( \frac{U(\Phi,\bar\Phi)}{T^4}\right)
+ \frac{2 N_c}{T^3}\sum_{q=u,d,s} \int \frac{d^3 {\bf p}}{(2\pi)^3}
 \left(\frac{e^{-\beta E_q^{-}(p)}}{g_q^-(p)} +  \frac{e^{-2\beta E_q^{+}(p)}}{g_q^+(p)}
\right) &= 0,\label{eq_Phi}\\
-\frac{d}{d \bar\Phi}\left( \frac{U(\Phi,\bar\Phi)}{T^4}\right)
+ \frac{2 N_c}{T^3}\sum_{q=u,d,s}  \int \frac{d^3 {\bf p}}{(2\pi)^3}
 \left(\frac{e^{-\beta E_q^{+}(p)}}{g_q^+(p)} +  \frac{e^{-2\beta E_q^{-}(p)}}{g_q^-(p)}
\right) &= 0,\label{eq_Phibar}\\
m_0^2 \phi_N + \left(\lambda_1 + \frac{1}{2} \lambda_2 \right)
\phi_N^3 + \lambda_1 \phi_N \phi_S^2 - h_N
+\frac{g_F}{2}N_c\left(\langle u{\bar u}\rangle_{_{T}} + \langle d{\bar
  d}\rangle_{_{T}} \right) &= 0,\label{eq_phiN}\\
m_0^2 \phi_S + \left(\lambda_1 + \lambda_2 \right)
\phi_S^3 + \lambda_1 \phi_N^2 \phi_S - h_S
+\frac{g_F}{\sqrt{2}}N_c \langle s{\bar s}\rangle_{_{T}} &= 0,\label{eq_phiS}
\end{align}
where
\begin{align*}
g_q^+(p) &= 1 + 3\left( \bar\Phi + \Phi e^{-\beta E_q^{+}(p)} \right)
e^{-\beta E_q^{+}(p)} + e^{-3\beta E_q^{+}(p)}, \\
g_q^-(p) &= 1 + 3\left( \Phi + \bar\Phi e^{-\beta E_q^{-}(p)} \right)
e^{-\beta E_q^{-}(p)} + e^{-3\beta E_q^{-}(p)}, \\
E_q^{\pm}(p) =& E_q(p) \mp \mu_B/3,\; E_{u/d}(p) = \sqrt{p^2 +
  m_{u/d}^2},\; E_{s}(p) = \sqrt{p^2 + m_{s}^2}, \\
\end{align*}
and
\begin{align}
\langle q{\bar q}\rangle_{_{T}} &= -4m_q \int \frac{d^3
  {\bf p}}{(2\pi)^3}\frac{1}{2E_q(p)}\left(1 - f^-_\Phi(E_q(p)) -
  f^+_\Phi(E_q(p))\right),
\end{align}
with the modified distribution functions
\begin{align}
f^+_\Phi(E_p) & =\frac{ \left( \bar\Phi + 2\Phi e^{-\beta\left(E_p-\mu_q\right)} \right) e^{-\beta\left(E_p-\mu_q\right)} +
  e^{-3\beta\left(E_p-\mu_q\right)} } {1 + 3\left( \bar\Phi + \Phi e^{-\beta\left(E_p-\mu_q\right)} \right) e^{-\beta\left(E_p-\mu_q\right)} +
  e^{-3\beta\left(E_p-\mu_q\right)}}, \nonumber\\
f^-_\Phi(E_p) & =\frac{ \left( \Phi + 2 \bar\Phi e^{-\beta\left(E_p+\mu_q\right)} \right) e^{-\beta\left(E_p+\mu_q\right)} +
  e^{-3\beta\left(E_p+\mu_q\right)} }{1 + 3\left( \Phi + \bar\Phi e^{-\beta\left(E_p+\mu_q\right)} \right) e^{-\beta\left(E_p+\mu_q\right)} +
  e^{-3\beta\left(E_p+\mu_q\right)}}. \nonumber
\end{align}

\section{Results}
\label{sec-res}

\begin{figure}[htb]
  \centering
  \begin{minipage}{.48\textwidth}
  \centering
  \includegraphics[width=1.1\textwidth]{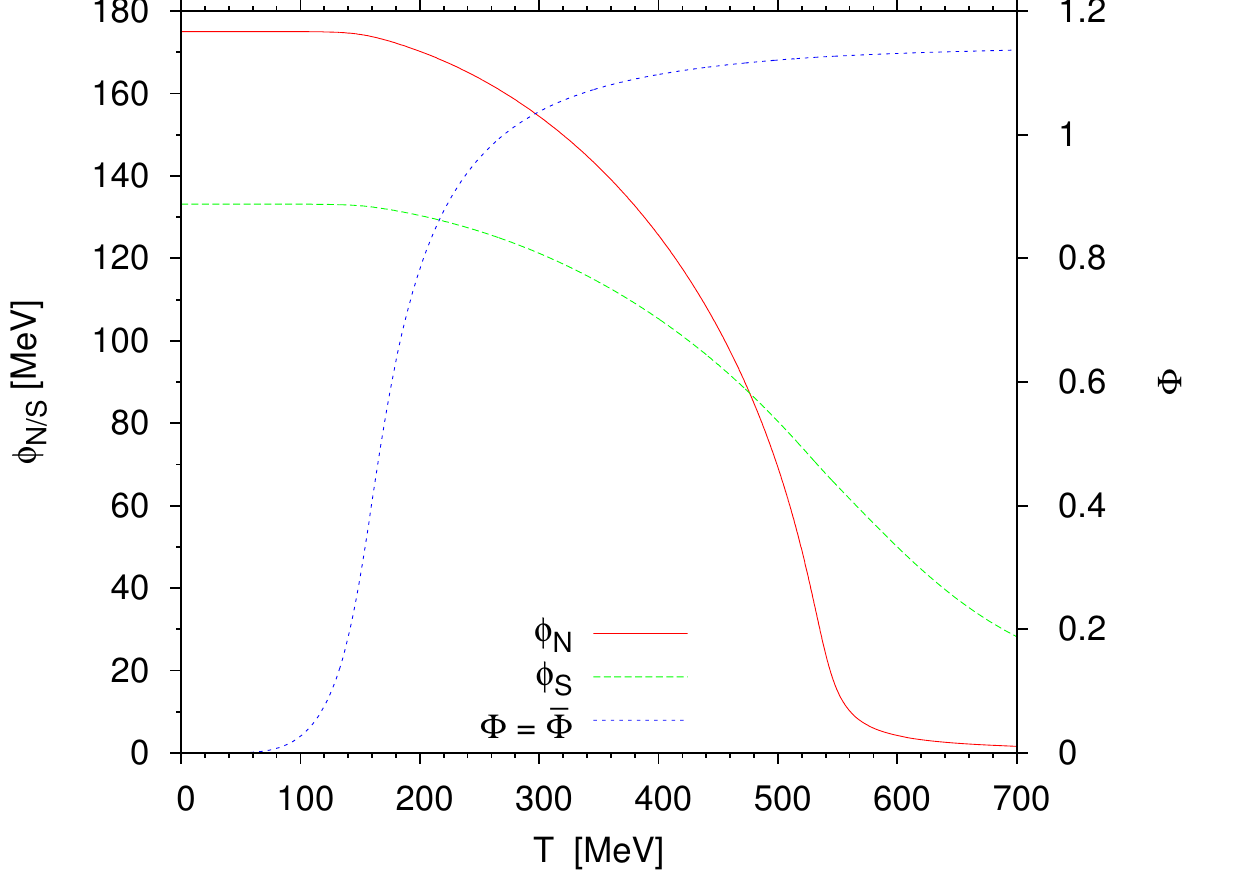}
  \caption{Temperature dependence of the order parameters with
    $m_{\sigma}=1.3$ GeV}
  \label{fig-fields_high_sigma}
\end{minipage}
\hspace*{0.02\textwidth}
\begin{minipage}{.48\textwidth}
  \centering
  \includegraphics[width=1.1\textwidth]{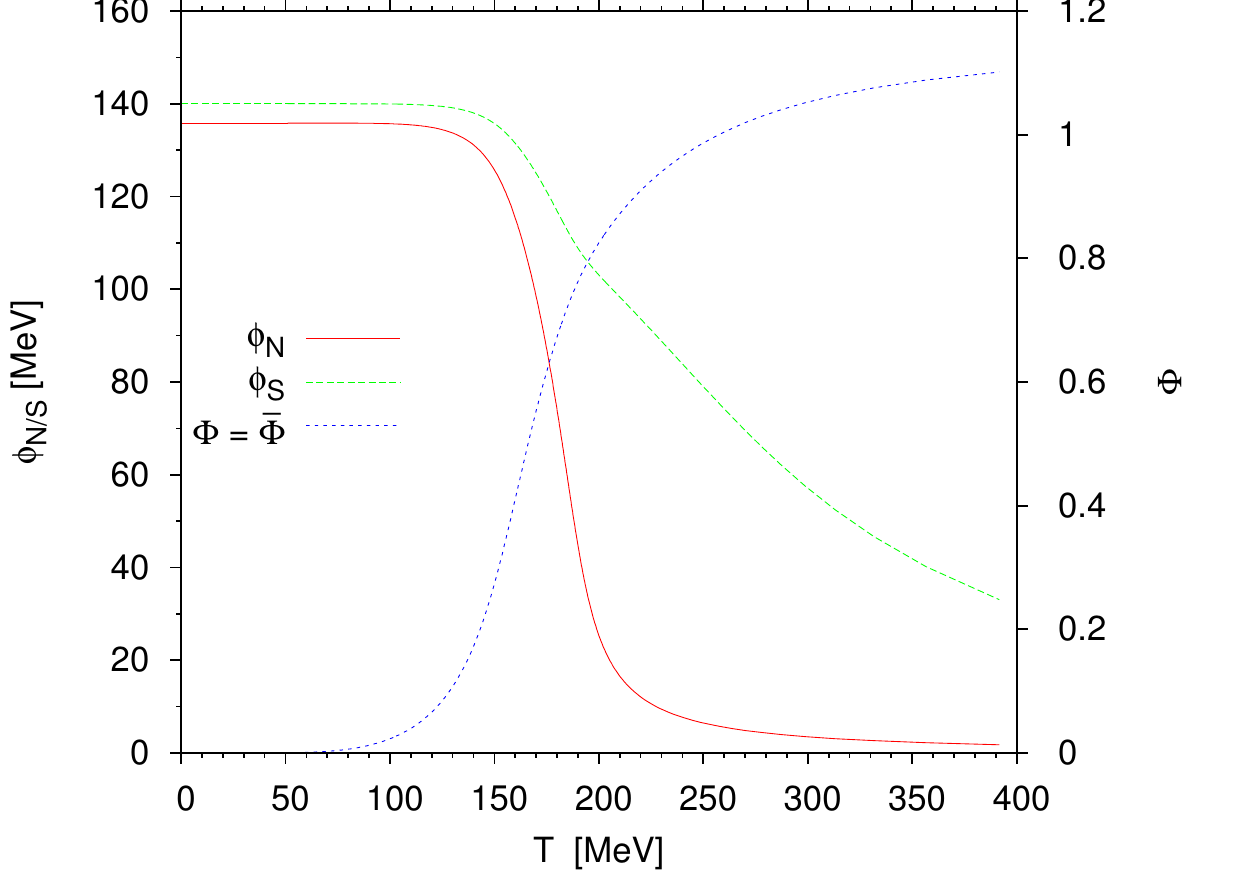}
  \caption{Temperature dependence of the order parameters with
    $m_{\sigma}=0.4$ GeV}
  \label{fig-fields_low_sigma}
\end{minipage}
\end{figure}

Solving Eqs.~\ref{eq_Phi}-\ref{eq_phiS} we get the temperature
dependence of the order parameters, which can be seen in
Fig.~\ref{fig-fields_high_sigma}. 
In \cite{Parganlija_2013} it was shown that the $q\bar{q}$ scalar nonet most 
probably contains $f_0$'s with masses higher than $1$~GeV. If we set 
$\lambda_1=0$ we get $m_{f_0^L} = 1.3$~GeV, which is in agreement with 
\cite{Parganlija_2013}. However in this case we get a very high pseudocritical 
temperature, $T_c\approx 550$~MeV, for $\phi_N$, which is much larger than 
earlier results (e.g. on lattice $T_c\approx 150$~MeV \cite{Aoki}). Now, if we 
tune $\lambda_1$ to get $m_{f_0^L} =400$ MeV (which corresponds to the physical 
particle $f_0(500)$), than $T_c$ goes down to $150-200$~MeV, which can be
seen in Fig.~\ref{fig-fields_low_sigma}. This finding is in line with the 
results of \cite{SchaeferWagner}, where they used a similar model, but without 
vector mesons. This suggests that in order
to get a good pseudocritical temperature we would need a
scalar-isoscalar particle with low mass ($\sim 400$~MeV),
which is probably not a $q\bar{q}$ state according to
\cite{Parganlija_2013}.
\begin{figure}[htb]
  \centering
  \begin{minipage}{.48\textwidth}
  \centering
  \includegraphics[width=1.1\textwidth]{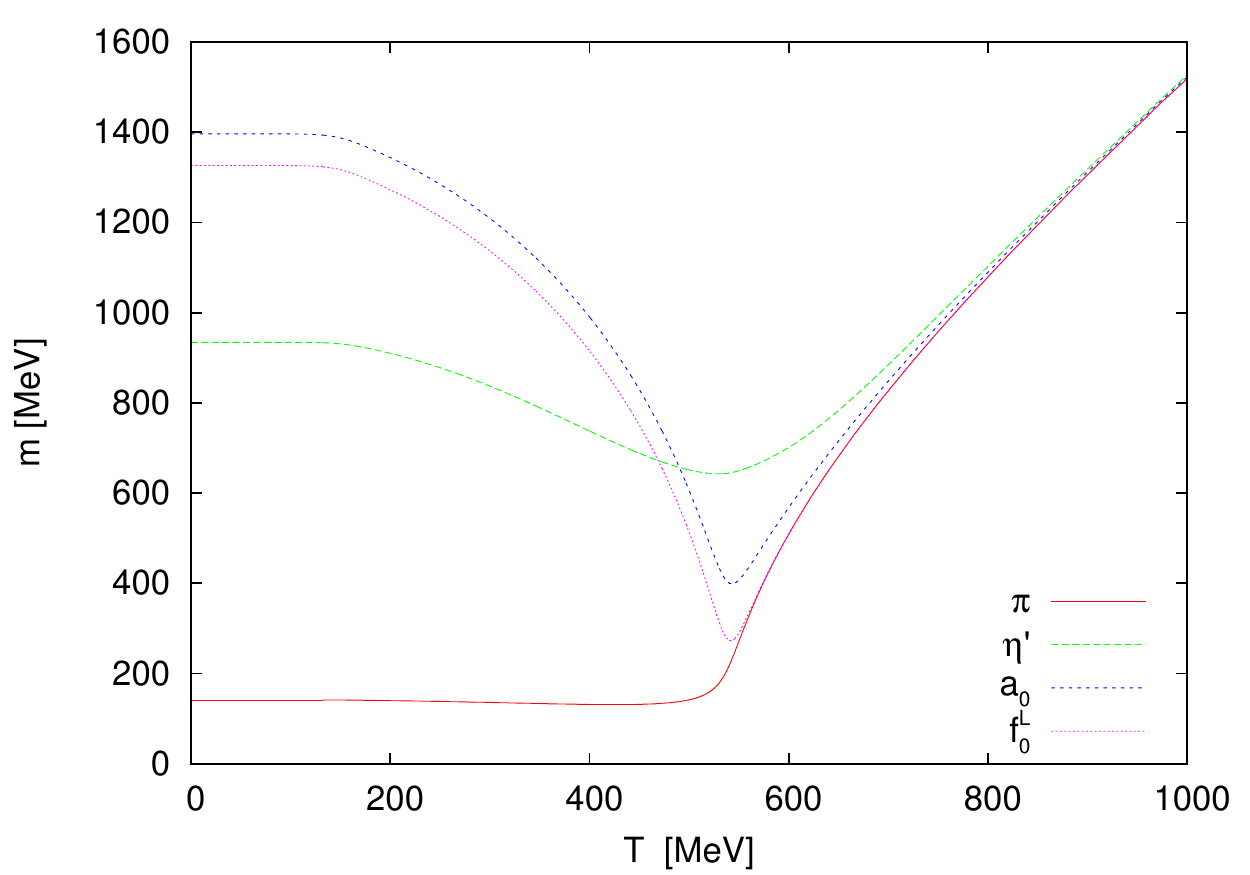}
  \caption{Temperature dependence of the scalars with $m_{\sigma}=1.3$ GeV}
  \label{fig-mass_high_sigma}
\end{minipage}
\hspace*{0.02\textwidth}
\begin{minipage}{.48\textwidth}
  \centering
  \includegraphics[width=1.1\textwidth]{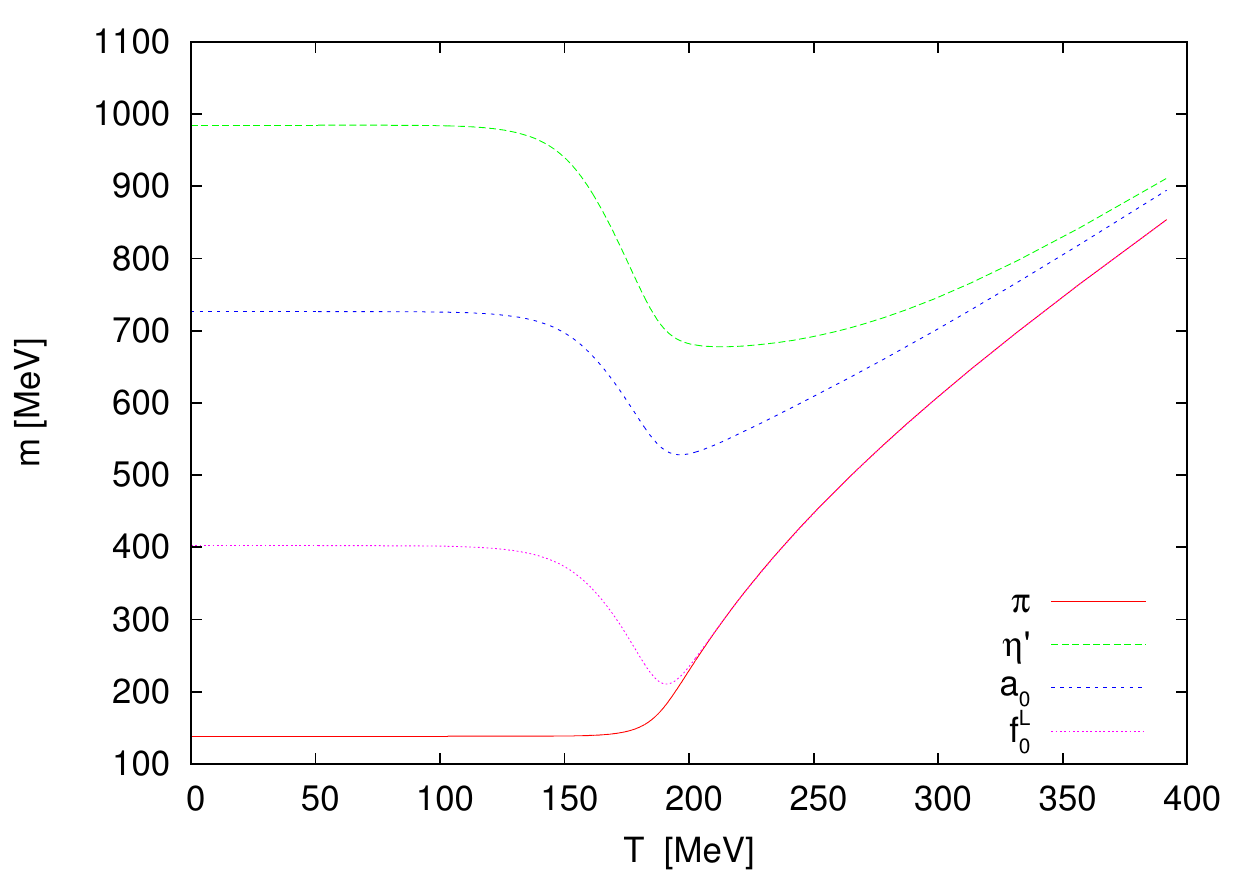}
  \caption{Temperature dependence of the scalars with $m_{\sigma}=0.4$ GeV}
  \label{fig-mass_low_sigma}
\end{minipage}
\end{figure}
In Fig. \ref{fig-mass_high_sigma} and \ref{fig-mass_low_sigma} we show the 
temperature dependence of the scalar meson masses. The mass of the parity partners ($\pi$ and $f_0^L$) reaches the same value above the phase transition temperature.

\section{Acknowledgement}
Authors were supported by the Hungarian OTKA fund K109462 and by the HIC for 
FAIR Guest Funds of the Goethe University Frankfurt.


\end{document}